\documentclass[prd,floats,preprint,superscriptaddress,eqsecnum,floatfix,nofootinbib,12pt]{revtex4} 
\pdfoutput=1 
\usepackage{amsmath} 
\usepackage{graphics, graphicx} 
\usepackage{epsfig} 
\usepackage{ulem} 
\usepackage{color} 

\def\1p{{(1p)}}
\def\be{\begin{equation}}
\def\ee{\end{equation}}
\def\beq{\begin{eqnarray}}
\def\eeq{\end{eqnarray}}
\def\p0{\phi_0}
\def\z0{\zeta_0}

\def\3G{^3{\cal G}}

\def\ol2{\frac{1}{\ell^2}}

\newcommand{\ttle}[1]{{\it #1}}

\begin{document}

\vspace{1cm}

\title{Information Preservation and Weather Forecasting for Black Holes\footnote{Talk given at the fuzz or fire workshop, The Kavli Institute for Theoretical Physics, Santa Barbara, August 2013}}

\author{S. W. Hawking}
\affiliation{DAMTP, University of Cambridge, UK}

\bibliographystyle{unsrt}

\vspace{1cm}

\begin{abstract}
It has been suggested [1] that the resolution of the information paradox for evaporating black holes is that the holes are surrounded by firewalls, bolts of outgoing radiation that would destroy any infalling observer. Such firewalls would break the CPT invariance of quantum gravity and seem to be ruled out on other grounds. A different resolution of the paradox is proposed, namely that gravitational collapse produces apparent horizons but no event horizons behind which information is lost. This proposal is supported by ADS-CFT and is the only resolution of the paradox compatible with CPT. The collapse to form a black hole will in general be chaotic and the dual CFT on the boundary of ADS will be turbulent. Thus, like weather forecasting on Earth, information will effectively be lost, although there would be no loss of unitarity. 
\end{abstract}

\maketitle

Some time ago [2] I wrote a paper that started a controversy that has lasted until the present day. In the paper I pointed out that if there were an event horizon, the outgoing state would be mixed. If the black hole evaporated completely without leaving a remnant, as most people believe and would be required by CPT, one would have a transition from an initial pure state to a mixed final state and a loss of unitarity. On the other hand, the ADS-CFT correspondence indicates that the 7evaporating black hole is dual to a unitary conformal field theory on the boundary of ADS. This is the information paradox. 

Recently there has been renewed interest in the information paradox [1]. The authors of [1] suggested that the most conservative resolution of the information paradox would be that an infalling observer would encounter a firewall of outgoing radiation at the horizon. 

There are several objections to the firewall proposal. First, if the firewall were located at the event horizon, the position of the event horizon is not locally determined but is a function of the future of the spacetime.

Another objection is that calculations of the regularized energy momentum tensor of matter fields are regular on the extended Schwarzschild background in the Hartle-Hawking state [3, 4]. The outgoing radiating Unruh state differs from the Hartle-Hawking state in that it has no incoming radiation at infinity. To get the energy momentum tensor in the Unruh state one therefore has to subtract the energy momentum tensor of the ingoing radiation from the energy momentum in the Hartle-Hawking state.  The energy momentum tensor of the ingoing radiation is singular on the past horizon but is regular on the future horizon. Thus the energy momentum tensor is regular on the horizon in the Unruh state. So no firewalls. 

For a third objection to firewalls I shall assume that if firewalls form around black holes in asymptotically flat space, then they should also form around black holes in asymptotically anti deSitter space for very small lambda. One would expect that quantum gravity should be CPT invariant. Consider a gedanken experiment in which Lorentzian asymptotically anti deSitter space has matter fields excited in certain modes. This is like the old discussions of a black hole in a box [5]. Non-linearities in the coupled matter and gravitational field equations will lead to the formation of a black hole [6]. If the mass of the asymptotically anti deSitter space is above the Hawking-Page mass [7], a black hole with radiation will be the most common configuration. If the space is below that mass the most likely configuration is pure radiation. 

Whether or not the mass of the anti deSitter space is above the Hawking-Page mass the space will occasionally change to the other configuration, that is the black hole above the Hawking-Page mass will occasionally evaporate to pure radiation, or pure radiation will condense into a black hole. By CPT the time reverse will be the CP conjugate. This shows that, in this situation, the evaporation of a black hole is the time reverse of its formation (modulo CP), though the conventional descriptions are very different. Thus if one assume quantum gravity is CPT invariant, one rules out remnants, event horizons, and firewalls.

Further evidence against firewalls comes from considering asymptotically anti deSitter to the metrics that fit in an S1 cross S2 boundary at infinity. There are two such metrics: periodically identified anti deSitter space, and Schwarzschild anti deSitter. Only periodically identified anti deSitter space contributes to the boundary to boundary correlation functions because the correlation functions from the Schwarzschild anti deSitter metric decay exponentially with real time [8, 9]. I take this as indicating that the topologically trivial periodically identified anti deSitter metric is the metric that interpolates between collapse to a black hole and evaporation. There would be no event horizons and no firewalls. 

The absence of event horizons mean that there are no black holes - in the sense of regimes from which light can't escape to infinity. There are however apparent horizons which persist for a period of time. This suggests that black holes should be redefined as metastable bound states of the gravitational field. It will also mean that the CFT on the boundary of anti deSitter space will be dual to the whole anti deSitter space, and not merely the region outside the horizon. 

The no hair theorems imply that in a gravitational collapse the space outside the event horizon will approach the metric of a Kerr solution. However inside the event horizon, the metric and matter fields will be classically chaotic. It is the approximation of this chaotic metric by a smooth Kerr metric that is responsible for the information loss in gravitational collapse. The chaotic collapsed object will radiate deterministically but chaotically. It will be like weather forecasting on Earth. That is unitary, but chaotic, so there is effective information loss. One can't predict the weather more than a few days in advance.

\centerline{\Huge}

\end{document}